\documentclass[
reprint,
superscriptaddress,
amsmath,amssymb,
aps,
pra,
longbibliography,
]{revtex4-2}

\usepackage{Packages}

\newcommand{\affilUKON}{Department of Physics, University of Konstanz, 78457 Konstanz, Germany.}
\newcommand{\affilETHZ}{Institute for Theoretical Physics, ETH Zurich, 8093 Zurich, Switzerland.}

\usepackage{tikz}
\usepackage{lmodern}
\newcommand\encircle[1]{%
  \protect\tikz[baseline=(X.base)] 
    \protect\node (X) [draw, shape=circle, inner sep=-1.2pt, scale=0.9] {\strut \scriptsize #1};}

\begin{document}
\preprint{APS/123-QED}

\title{Fluctuation instabilities via internal resonance in a multimode membrane as a mechanism for frequency combs}

\author{Mengqi Fu}
\thanks{These authors contributed equally to this work.}
\affiliation{\affilUKON}

\author{Orjan Ameye}
\thanks{These authors contributed equally to this work.}
\affiliation{\affilUKON}

\author{Fan Yang}\affiliation{\affilUKON}
\author{Jan Ko\v{s}ata}\affiliation{\affilETHZ}
\author{Javier del Pino}\affiliation{\affilUKON}

\author{Oded Zilberberg}
\email{oded.zilberberg@uni-konstanz.de}
\affiliation{\affilUKON}
\author{Elke Scheer}
\email{elke.scheer@uni-konstanz.de}
\affiliation{\affilUKON}

\date{\today}

\begin{abstract}
We explore self-induced parametric coupling, also called internal resonances (IRs), in a membrane nanoelectromechanical system. Specifically, we focus on the formation of a limit cycle manifesting as a phononic frequency comb. 
Utilizing a pump-noisy-probe technique and theoretical modeling, we reveal the behavior of mechanical excitations revealing themselves as sidebands of the stationary IR response. We find that when the energy-absorbing excitation of a lower mode is parametrically-upconverted to hybridize with a higher mode, significant squeezing and bimodality in the upper mode occurs. Instead, when the upconverted absorbing excitation hybridizes with an emitting sideband of the higher mode, a Hopf bifurcation occurs and a limit cycle forms, manifesting as a frequency comb.
We thus reveal a unique mechanism to obtain frequency combs in parametrically-coupled modes. We furthermore demonstrate a rich variety of IR effects, the origin of which significantly extends beyond standard linear parametric coupling phenomena. Our findings enhance the understanding of energy transfer mechanisms with implications for advanced
sensing technologies and novel phononic metamaterials.
\end{abstract}
\maketitle

Nonlinear systems are prevalent across physical, biological, and engineering domains. They exhibit rich phenomena, including bifurcations, self-sustained limit cycles (LCs), and chaos~\cite{Strogatz2000,Edelstein_book}. Recently, there is growing interest in nonlinear physics driven by the advancements in the fabrication and study of micro- and nanoelectromechanical systems (MEMS and NEMS). Their dynamics are prominently nonlinear, a consequence of surface forces overtaking volume forces at small length scales~\cite{Lifshitz_Cross,poot2012}. Despite their different origins, similar nonlinear effects across fields like optics, polariton physics, superconducting circuits, and fluid dynamics share a common theoretical underpinning~\cite{nayfeh2008nonlinear,boyd2012contemporary,eichler2023classical}.

Unlike linear systems, where the normal modes are independent, nonlinear systems allow them to interact and exchange energy. When two normal modes meet internal resonance (IR) conditions, with frequencies approaching specific integer ratios, energy exchange becomes resonant~\cite{guckenheimer_1990}. These interactions often involve linear parametric coupling mediated by three-wave or four-wave mixing~\cite{Frimmer_2014,HalgStrong2022}, with an external drive enhancing the coupling strength and enabling dynamic control of interactions. This functionality benefits applications ranging from advanced sensors and energy transfer~\cite{kamoto2013,Kosata_2020} to the creation of artificial metamaterials with unique properties, e.g. nontrivial topologies~\cite{xu2016topological,YuanSynthetic2018, TomokiSynthetic2016, delPino2022NH,BusnainaQuantum2024}. Parametric coupling also plays a crucial role in quantum metrology and information processing, serving as a key resource for generating two-mode squeezing~\cite{boyd2012contemporary}. 

Nonlinear dynamics lead to multiple non-equilibrium stationary solutions (NESSs) and associated phase space topologies with bifurcations driving phase transitions~\cite{dumont2024hamiltonian,villa2024MStop}. The behavior of NESSs is echoed in the dynamics of excitations around fixed points, with telltales of squeezed fluctuations alongside, overdamped-to-underdamped transitions and instabilities~\cite{eichler2023classical}. Monitoring these excitations through pump-noisy-probe (PNP) spectroscopy provides insights into NESS behavior~\cite{huber2020spectral, yang2021mechanically,delPino2022NH,heugel2023role}. Nonlinear coupling can also induce limit cycles (LCs), a type of NESS marked by self-sustained orbits with non-commensurate emerging frequencies~\cite{houri2019limit,Zambon2020,ganesan2018excitation,li2022nonlinear,delpino2023limit}. The LCs are often heralded by Hopf bifurcations, where the excitations become unstable via gain in the system. The prevalence of LCs in various fields such as neural rhythms, fluid flow transitions, disease outbreaks, business cycles, and pattern formation~\cite{strganac2000identification, wang2005epidemic, lorenz1993nonlinear, patternformationRMP}, makes it crucial to study the behavior of excitations around LCs.

In this work, we demonstrate how IR in nonlinearly coupled normal modes, leads to a LC and the emergence of a phononic frequency comb. The LC is sustained by the interplay of nonlinear coupling, external drive, and dissipation. Using PNP, we observe that the LC formation aligns with the closure of an excitation gap via resonant four-wave mixing. Interestingly, the outcome depends on the type of excitations that hybridize when the gap closes, i.e., whether an absorbing sideband resonates with an absorbing or emitting sideband. We thus reveal a unique parametric coupling mechanism for the formation of LCs, which mimicks the onset of ultra-strong coupling seen in light-matter systems~\cite{ultrastrong_coupling_RMP}.  Our results divulge a general mechanism that carries out beyond the realm of NEMS. 

\begin{figure}[t!]
  \includegraphics[width=\columnwidth]{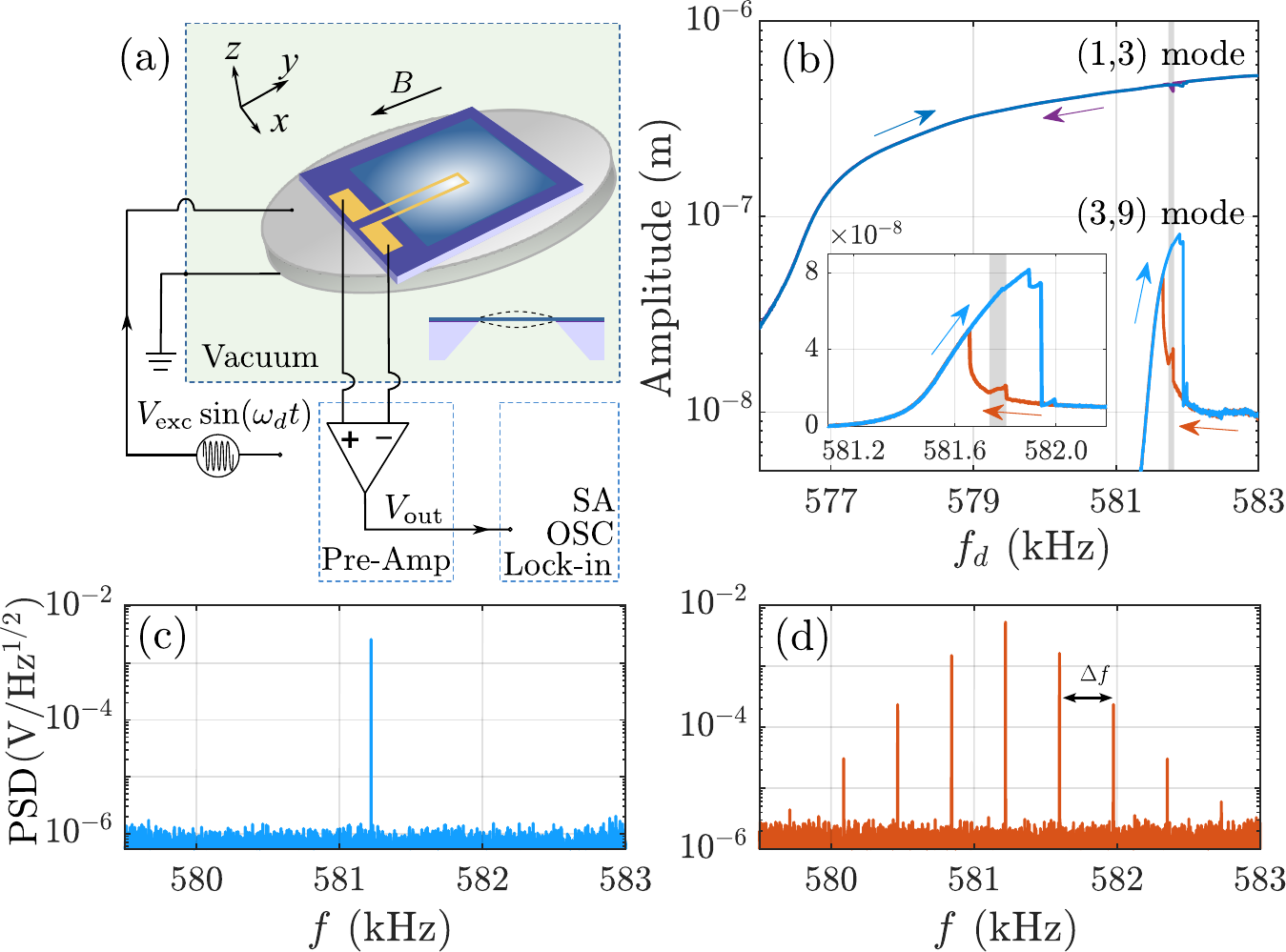}
  \caption{Setup and its NESS. (a) Schematic drawing of the SiN membrane resonator (cross section in the inset) and the on-chip detection scheme utilizing an inductive probe scheme with a detection electrode patterned on the surface of the membrane. The vibrating membrane resonator generates an induction voltage within a uniform magnetic field, with the output signal being proportional to the vibrational amplitude beneath the electrode. 
  (b) Amplitude response of the membrane resonator at the drive frequency $f_d$ and its 3rd harmonic $3f_d$, as $f_d$ is swept upward (blue arrows) and then downward (purple and orange arrows). The gray line marks the position where (c) and (d) are measured. Inset: Zoom-in on the response of the (3,9) mode ($m_2$). (c) PSD  around the $f_d$ peak along the upward sweep. (d) Same as (c) along the down sweep, marking the formation of a frequency comb with a frequency space $\Delta f$. All measurements were performed in vacuum and at room temperature. Here: $V_1$ = 100 mV.
  }
  \label{fig:setupandfrequencysweep}
\end{figure}

Our experimental setup, depicted in Fig.~\ref{fig:setupandfrequencysweep}(a), is built upon a suspended almost square-shaped silicon nitride (SiN) membrane framed by a silicon chip that is attached to a piezo disk. It supports kHz-frequency flexural vibration modes labeled with wavenumbers $(n,m)$, where $n$ and $m$ count the deflection extremes along the $x$- and $y$-directions. To excite these modes, a single harmonic alternating current (AC) electrical voltage  $V_1(t) = V_1\sin(\omega_d t)$ is applied across the piezo at a frequency $f_d\equiv \omega_d /2\pi$ near the eigenfrequency $f_1=\Omega_1/(2\pi)=576$ kHz of the $(1,3)$ mode; called $m_1$ henceforth.
The membrane vibrates along the $z$-axis perpendicular to an externally applied magnetic field, see Fig.~\ref{fig:setupandfrequencysweep}(a). This setup linearly transduces mechanical motion into readout voltage signals via electromagnetic induction, see Fig.~\ref{fig:setupandfrequencysweep}(b). 
Notably, we observe motion not only in the driven mode $m_1$ but also in the $(3,9)$ mode, denoted $m_2$ henceforth, with eigenfrequency \(f_2 =\Omega_2/2\pi = 1.738~\text{MHz}\). Note that \(f_2\approx 3f_1\), indicative of an IR. For further details on sample fabrication, setup, and data acquisition we refer to the Appendix
and Refs.~\cite{yang2021mechanically,yang2023quantitative,waitz2012mode,waitz2015spatially}.

Figure~\ref{fig:setupandfrequencysweep}(b) shows the amplitude response at the drive frequency $f_d$ and its 3$^{\rm rd}$ harmonic for up and down frequency sweeps. The  amplitude of $m_1$ follows along the high-amplitude branch of a Duffing-shaped response. Crucially, we observe that $m_2$ is internally driven by $m_1$'s high-amplitude motion, which induces a Duffing-shaped response in $m_2$: both its high and low amplitude branches appear along a hysteretic response relative to the up and down sweeps. We confirm that the observed IR stems from nonlinear coupling between $m_1$ and $m_2$ through ringdown measurements that exhibit a non-exponential decay of the intensity, see the Appendix
We also measure the power spectral density (PSD) along the sweep, see Figs.~\ref{fig:setupandfrequencysweep}(c) and (d). As expected, the PSD around the high-amplitude branch displays a single peak at $f_d$. Strikingly, during the down sweep along the low-amplitude branch, the PSD suddenly exhibits a frequency comb around $f_d$ with a spacing of approximately $\Delta f = 380$ Hz between adjacent sidebands. This spacing does not result from simple algebraic combinations or harmonics of the system's eigenfrequencies, suggesting a mechanism more complex than resonant wave mixing~\cite{seitner2017parametric}. The formation of a frequency comb in similar IR scenarios has been reported in several previous studies~\cite{houri2019limit,Zambon2020,ganesan2018excitation,li2022nonlinear}. 

\begin{figure*}[t!]
\includegraphics[width=\linewidth]{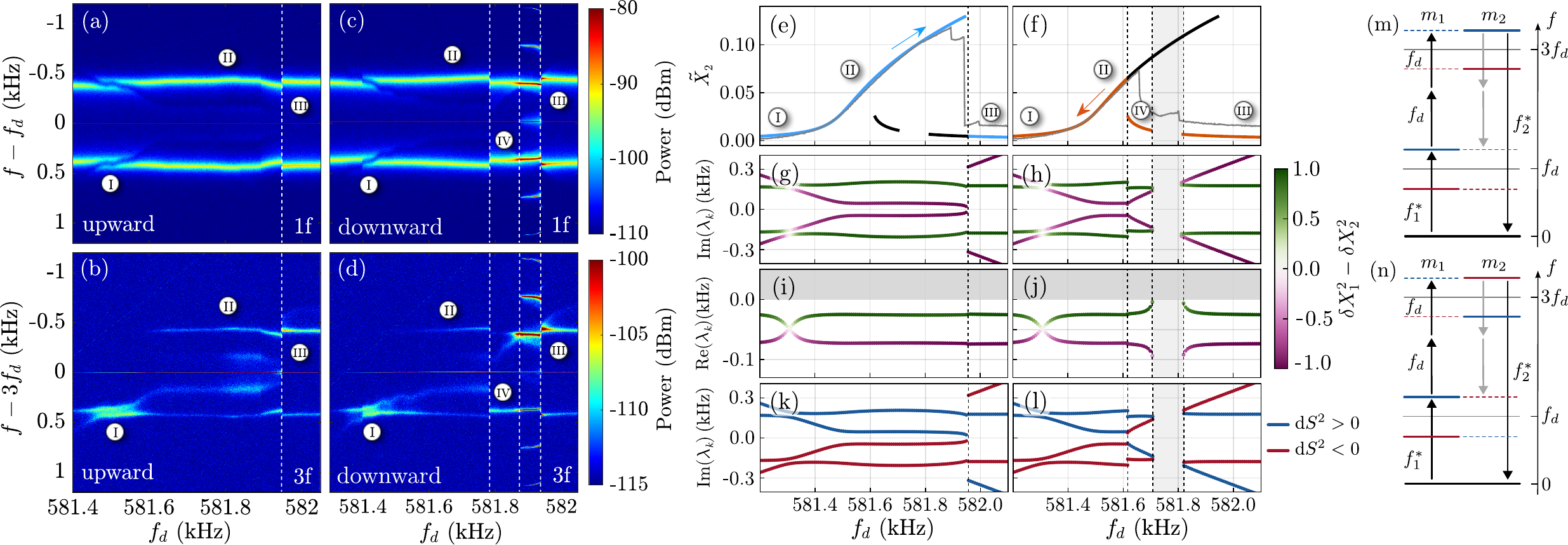}
  \caption{
  PNP response. (a) and (b) Sidebands around $f_d$ and $3f_d$ along an upsweep of $f_d$. (c) and (d) same as before for a downsweep of $f_d$. Dashed vertical lines indicate bifurcations. Significant phenomena in the response are labeled by markers \encircle{I} - \encircle{IV}. (e) and (f) Theory-predicted stationary amplitude response [cf.~Eqs.~\eqref{eq: EOM 1} and~~\eqref{eq: EOM 2}] of $m_2$ for the up- and downsweep, respectively. Experimentally sampled solutions appear in color and others in black. The experimental amplitude response is shown in gray. (g)-(l) Corresponding excitation eigenvalues on top of the sampled stationary solution [cf.~Eq.~\eqref{eq: linear_response}]. (g) and (h) Imaginary part of the eigenvalues colored by the state's weight in $m_1$ relative to $m_2$. (i) and (j) Real part of the eigenvalues with same coloring as (g) and (h). Panels (k) and (l) correspond to (g) and (h), but colored by the symplectic norm of the eigenstates. (m) Lab-frame resonant four-wave mixing diagram, coupling absorbing the sidebands in \encircle{I}, where $f_1^*=f_d+\mathrm{Im}(\lambda_1^>)$ and $f_2^*=f_d+\mathrm{Im}(\lambda_2^>)$ with $\mathrm{Im}(\lambda_j^{dS^2})$ the rotating-frame frequency of $m_j$ at sideband $dS^2\lessgtr0$. (n) Same as (m) with $f_2^*=f_d+\mathrm{Im}(\lambda_2^<)$ for the merging of an absorbing with an emitting sideband engendering the LC formation.
  }
  \label{fig: Fig 2}
\end{figure*}

We now apply a PNP scheme~\cite{huber2020spectral, yang2021mechanically,heugel2023role} to link the system's linear response with the frequency comb formation. The PNP involves driving the system to a NESS using a drive as before, complemented by a weak white-noise drive, with root mean square voltage \( V_\textrm{rms} = 100 \, \text{mV} \) and bandwidth 2.5 MHz. The latter probes small excitations around the NESS~\cite{yang2021mechanically}. The obtained PNP spectra around \(f_j\in(f_d,3f_d)\) are presented in Fig.~\ref{fig: Fig 2} for an up sweep [(a) and (b)] and a down sweep [(c) and (d)]. The PNP shows up to four peaks around each $f_j$, two at blue and two at red detuning. As $f_d$ is swept upward, we \encircle{I} observe an avoided crossing between the sidebands, followed by \encircle{II} increased visibility of all sidebands. A jump \encircle{II}~$\rightarrow$~\encircle{III} in the PNP response marks the transition between the high- and low-amplitude NESSs in $m_2$. Along the down sweep, we observe transitions between \encircle{III}~$\rightarrow$~\encircle{IV}, via the frequency comb regime. Eventually, the jump \encircle{IV}~$\rightarrow$~\encircle{II} marks the transition from the low- to the high-amplitude NESS of $m_2$.

We explain the experimental features using linear response theory in combination with a harmonic balance (HB) ansatz~\cite{kovsata2022harmonicbalance}.  Starting from the membrane’s stress-strain equations~\cite{nayfeh2008nonlinear}, we truncate the model to describe the dynamics of the low-frequency flexural modes $m_1$ and $m_2$. Their amplitudes $x_1,x_2$ then obey (see Appendix):
\begingroup
    \medmuskip=0.3mu 
    \begin{align} \label{eq: EOM 1}
    \ddot{x}_1&+\tilde{\Omega}_{1}^2(x_1) x_1+\tilde{\Gamma}_1(x_1) \dot{x}_1+3 \xi x_1^2 x_2+\zeta x_1 x_2^2=F_1(t)\,, \\
    \label{eq: EOM 2}
    \ddot{x}_2&+\tilde{\Omega}_{2}^2(x_2) x_2+\tilde{\Gamma}_2(x_2)\dot{x}_2+ \xi x_1^3+\zeta x_1^2 x_2=0\,.
\end{align}
\endgroup
Here, $\tilde{\Omega}_{j}(x_j)$ and $\tilde{\Gamma}_j(x_j)$ represent the effective frequencies and damping rates for mode $j$. Defined in terms of natural frequencies $\Omega_j$ and linear dampings $\Gamma_j$, they read $\tilde{\Omega}_{j}(x_j)=\sqrt{\Omega_j^2+\beta_j x_j^2}$ and $\tilde{\Gamma}_j(x_j)=\Gamma_j+\eta_j x_j^2$, where $\beta_{j}$ and $\eta_j$ denote the strengths of the intrinsic Duffing nonlinearity and the nonlinear damping coefficient, respectively. The coefficients $\xi$ and $\zeta$ in Eqs.~\eqref{eq: EOM 1} and \eqref{eq: EOM 2} quantify the strengths of the relevant four-wave mixing processes that couple the modes. The term $F_1(t)\propto V_1(t)$ represents the external force acting on $m_1$ as controlled by the applied AC voltage. Mechanical parameters are extracted by fitting their linear and weakly-nonlinear amplitude responses and are reported in the Appendix.

Assuming that each mode mostly responds at the drive frequency nearest its resonance frequency, we employ a multifrequency HB ansatz~\cite{nayfeh2008nonlinear, eichler2023classical,kovsata2022harmonicbalance}: 
\begin{align}
  x_1=&u_{1}\cos(\omega_d t)+v_{1}\sin(\omega_d t)\,,\label{eq:HBM_multi_1}\\
  x_2=&u_{2}\cos(3\omega_d t)+v_{2}\sin(3\omega_d t)\,,\label{eq:HBM_multi_2}
\end{align}
where \(\omega_d\) is close to \(\Omega_1\), making \(3\omega_d\) close to \(\Omega_2\).
Plugging the ansatz into Eqs.~\eqref{eq: EOM 1} and \eqref{eq: EOM 2} and averaging over the oscillation periods $2\pi/\omega_d$ and $2\pi/(3\omega_d)$, we obtain time-independent equations of motion for the slow amplitudes grouped into $\vb{u}$, in the form $\dot{\vb{u}}=F(\vb{u})$. Using HarmonicBalance.jl~\cite{kovsata2022harmonicbalance}, we automate this process and find the system's stationary solutions $\vb{u}^*$, where $F(\vb{u}^*)=0$, i.e., we find a plethora of NESSs as a function of $\omega_d$, see Figs.~\ref{fig: Fig 2}(e) and (f). Up- and downsweeps of $\omega_d$ sample the selected NESSs, yielding  good agreement with the experimentally observed stationary amplitudes $X_j=\sqrt{u_j^2+v_j^2}$.

To describe the PNP spectra, we derive linearized equations of motion around each NESS of Eqs.~\eqref{eq:HBM_multi_1}~and~\eqref{eq:HBM_multi_2}:
\begin{align} \label{eq: linear_response}
    \delta \dot{\vb{u}}_m= \mathcal{J}(\vb{u}^*_m) \, \delta \textbf{u}_m,
\end{align}
for the small deviations $\delta \vb{u}_m \equiv \vb{u} - \vb{u}^*_m$ around the $m^{\rm th}$ NESS  $\vb{u}^*_m$. Eigensolutions of Eq.~\eqref{eq: linear_response} take the form $ \mathbf{w}_k e^{\lambda_k t}$, with normal modes $\mathbf{w}_k=(c^{u_1}_{k},c^{v_1}_{k},c^{u_2}_{k},c^{v_2}_{k})$ and amplitudes $c^{l}_{k}$ and complex eigenvalues $\lambda_k \in \mathbb{C}$.
The eigenvalues $\lambda_k$ define an excitation characterized by a frequency $\mathrm{Im}(\lambda_k)$ and a lifetime $\mathrm{Re}(\lambda_k)$.
In  Figs.~\ref{fig: Fig 2}(g)-(j), we plot the obtained $\lambda_k$ for the respective NESS sampled in the experiment. We color the eigenvalues with the relative contribution of $m_1$ and $m_2$ to the corresponding eigenmode, i.e., using \( \delta X_k^1-\delta X_k^2 \) with \( \delta X_k^j = \sqrt{\left(c^{u_j}_{k}\right)^2+\left(c^{v_j}_{k}\right)^2} \). 
In  Figs.~\ref{fig: Fig 2}(k) and (l), we repeat (g) and (h) with a coloring based on the symplectic norm, $\dd{S}^2=i\sum_j\left[c^{u_j}_{k} (c^{v_j}_{k})^*-(c^{u_j}_{k})^* c^{v_j}_{k}\right]$, which quantifies whether the excitation tends to absorb ($\dd{S}^2>0$; particle-like) or release/emit energy ($\dd{S}^2<0$; hole-like). One can understand this distinction as a bare mode excitation rotating faster or slower than the respective rotating frame at frequency $\omega_d$ or $3\omega_d$~\cite{Soriente_2020,dumont2024hamiltonian}.
Since PNP probes all excitations in the lab frame, we rotate the solved excitations back to the lab frame using the inverse of Eqs.~\eqref{eq:HBM_multi_1} and \eqref{eq:HBM_multi_2}. This results in four spectral sidebands around both $\omega_d$ and $3\omega_d$, with an amplitude imbalance that encodes mode squeezing~\cite{Huber_2020}, and spectral widths that account for the lifetimes~\cite{Soriente_2020,Soriente_2021}, see Fig.~\ref{fig: Fig 2}.

We first apply this theory to the upsweep experiment. As the system climbs the high-amplitude Duffing state of $m_1$, the frequency of its respective excitation renormalizes due to a spring shift mechanism~\cite{aspelmeyer2014cavity}, analogously to an AC Stark shift~\cite{NBDelone_1999}.
In the strongly-driven regime, the renormalization saturates, and $m_1$'s excitation frequency flattens in the rotating frame~\cite{Steele2021ACStark}. Here, $m_2$'s excitation aligns with its bare frequency $\omega_2$; in the rotating frame, its eigenfrequency hence slopes linearly with detuning. When the two excitation frequencies approach, they exhibit an avoided crossing \encircle{I}.
This hybridization mirrors conventional parametric coupling between harmonic oscillators~\cite{faust2012nonadiabatic, mahboob2012phonon,Frimmer_2014,mathew2020, ma2021nanomechanical,HalgStrong2022}. Importantly, the parametric coupling here arises from the nonlinear wave mixing between $m_1$ and $m_2$ in Eqs.~\eqref{eq: EOM 1} and \eqref{eq: EOM 2}; when the upconverted absorbing sideband of $m_1$  nears \(\Omega_2\), the wave mixing is resonant, see Fig.~\ref{fig: Fig 2}(m).
Note that the avoided crossing occurs when the absorbing  
sidebands [cf.~Figs.~\ref{fig: Fig 2}(k),(l), and (m)] are resonant.

When detuned far above the avoided crossing [\encircle{II} in Fig.~\ref{fig: Fig 2}], the sideband response of $m_2$ also becomes mostly independent of the detuning, as both modes saturate at high amplitude.
In this region, the appearance of symmetric PNP sidebands around $\omega_d$ and $3\omega_d$ indicates squeezing due to four-wave mixing, growing symmetry in amplitudes reflects an increasing degree of squeezing~\cite{huber2020spectral,DykmanBook}.
At even higher detuning,
both sidebands of $m_2$ move toward $\mathrm{Im}(\lambda_k)=0$ in the rotating frame. This excitation mode softening signals that $m_2$'s sidebands become resonant with the upconverted pump at $3\omega_d$. As a consequence, the $m_2$ NESS becomes unstable and transitions from the high-amplitude to the low-amplitude NESS. 
After this transition into \encircle{III}, the system finds itself where the upward sweep started: $m_1$ is still in its high-amplitude Duffing state and $m_2$ is in its lower-amplitude state. However, the $m_2$ excitation is now lower than the upconverted pump in $3\omega_d>\Omega_2$, i.e., $dS^2$ of $m_2$ in \encircle{III} is inverted relative to \encircle{I}.

At the onset of the downsweep in \encircle{III}, the system mirrors the initial conditions of the upsweep experiment: $m_2$'s frequency remains linearly susceptible to the detuning, while $m_1$ shows squeezing signatures [cf. sideband symmetry in Fig.~\ref{fig: Fig 2}(c) and (d)], independent of the detuning. As the downsweep progresses, the modes' eigenfrequencies come closer. Unlike the mode hybridization in \encircle{I}, the NESS now destabilizes, leading to a LC that manifests as a frequency comb in the PSD [cf.~Fig.~\ref{fig: Fig 2}(c) and (d)]. This instability arises at a Hopf bifurcation, where the eigenmode's lifetime vanishes while retaining a non-zero frequency ($\mathrm{Im}(\lambda_k)\neq0$), as seen in Fig.~\ref{fig: Fig 2}(h) and (j). Contrary to the resonance behavior in \encircle{I}, the absorbing sideband of $m_1$ encounters the emitting one of $m_2$, which creates a feedback mechanism that destabilizes $m_1$ [cf.~Fig.~\ref{fig: Fig 2} (j) and (n)]. Thus, vibrations in $m_1$ are upconverted to $m_2$ via nonlinear parametric coupling, and reconverted back to the resonant sideband of $m_1$ due to ultrastrong coupling~\cite{HalgStrong2022}. This instability resembles that caused by mode gap closure due to ultrastrong parametric coupling~\cite{Aldana2013,Peterson2019,Zambon2020}. Hence, we reveal the microscopic origin of the LC formation in such systems, which is the main result of this work. Proceeding with the downsweep, the LC collapses back into \encircle{IV}, stabilizing both $m_1$ and $m_2$. 
When the sidebands of $m_2$ get close to resonance with the upconverted pump once more, $m_2$ enters its high-amplitude Duffing state, with a flip in the sign of $\dd{S}^2$.

\begin{figure}[t!]
  \includegraphics[width=1\columnwidth]{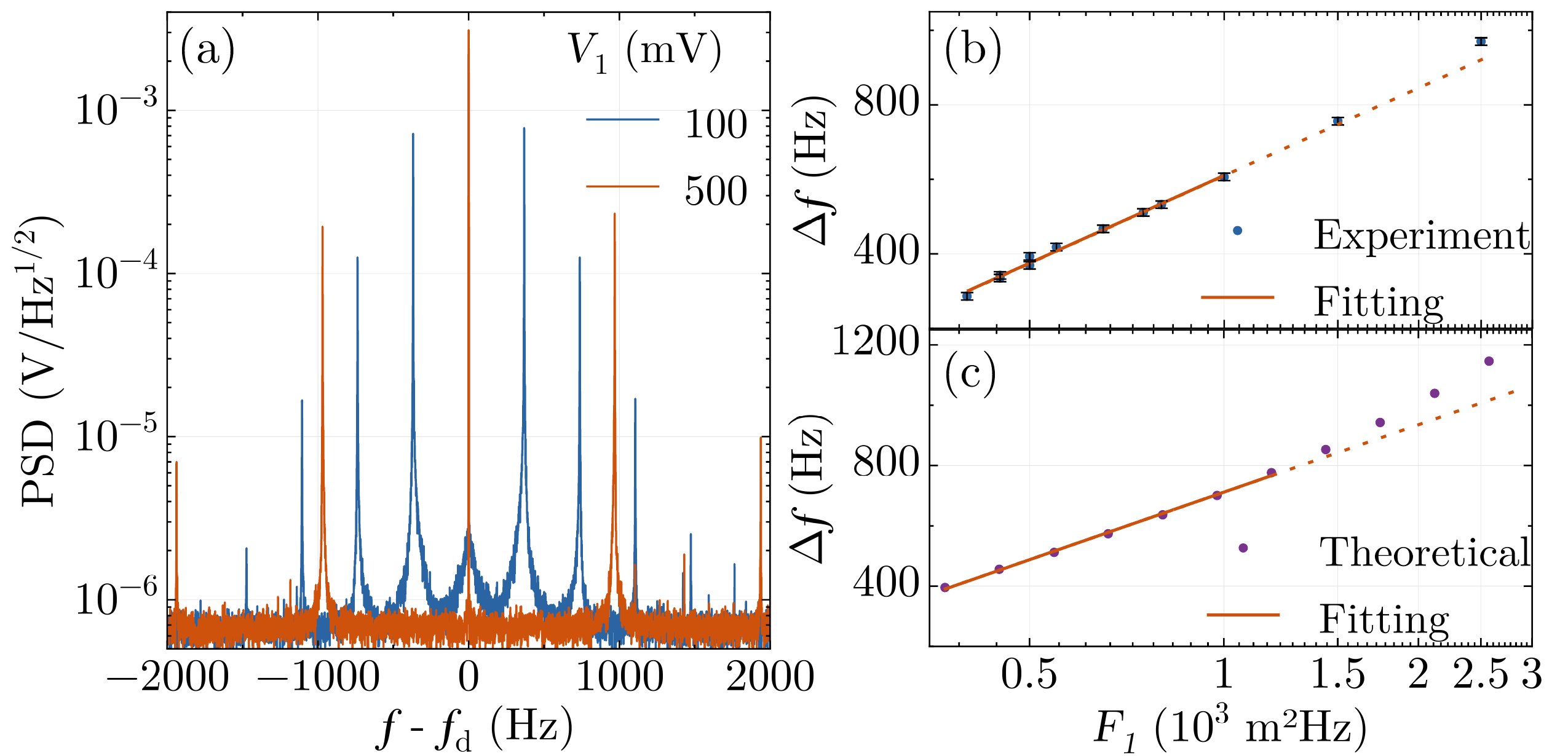}
  \caption{(a) PNP-measured frequency spectra of the frequency comb for $V_1$ = 100 and 500\,mV. The injected white noise has $V_\textrm{rms}$ = 100 mV and bandwidth 2.5 MHz. (b) measured and (c) predicted dependence of $\Delta f$ on $F_1$, respectively.
  }
  \label{fig:comb-variousVexc}
\end{figure}

The power intensity and bandwidth of the frequency combs can be tuned by adjusting $\omega_d$, while approximately maintaining their free spectral range (i.e., spacing $\delta\omega = 2\pi \Delta f$), as shown in Fig.~9 in the Appendix
During a downsweep, the PSD initially increases in the frequency range where the noise sidebands converge, see Figs.~\ref{fig: Fig 2}(c) and (d). The frequency comb remains stable across 
 a wide range of $V_1$, as shown in Fig.~\ref{fig:comb-variousVexc}(a) and Fig.~9 in the Appendix.
 The free spectral range of the comb is strongly dependent on the input power, with $\Delta \omega$ showing a logarithmic dependence on $F_1$ when $F_1 \sim 10^{3}~\mathrm{m}^{2}$Hz ($V_1 \sim 200$ mV), as seen in Fig.~\ref{fig:comb-variousVexc}(b). This behavior aligns with the theoretical prediction, where $\Delta f$ is derived from the fluctuations across the Hopf bifurcation (Fig.~\ref{fig:comb-variousVexc}(c)). This simple dependence of $\Delta f$ on $V_1$ makes the control of $\Delta f$ practical. As $F_1$ further increases, $\Delta f$ gradually deviates from the logarithmic relationship both in experiment and theory, see Fig.~\ref{fig:comb-variousVexc}(b) and (c). Beside, the intensity of the frequency comb decreases as $V_1$ increases. Fig.~\ref{fig:comb-variousVexc}(a) shows the frequency comb for $V_1$ = 100\,mV, where it shows the highest intensity, and for 500\,mV, where the intensity of the innermost sideband has decreased to about 1/4 compared to that for $V_1$ = 100\,mV.

In conclusion, our study opens avenues for exploring the fundamental physics of multimode resonator networks, with implications for metamaterial science, sensing technologies, and neuromorphic computing including quantum implementations.
We believe that the mechanism of ultrastrong coupling between sidebands as a source for generating LCs carries over beyond IR scenarios. Thus, we motivate the study of LCs via their fluctuation spectra in multimode systems. Such effects will manifest in  synchronization phenomena is synthetic dimension systems and harbors a unique approach to tunable parametric coupling. Future work will explore the potential of such LCs instabilities for ground state cooling.

\acknowledgments
The authors thank J. Boneberg for help with the experimental setups. We are indebted to J. S. Ochs, E. M. Weig, G. Rastelli, P. Leiderer and W. Belzig for fruitful discussion and comments about the work. The authors acknowledge the use of the experimental equipment and the expert support provided by the nano.lab at the University of Konstanz. The authors gratefully acknowledge financial support from the China Scholarship Council, the A. v. Humboldt Foundation, the Swiss National Science Foundation (SNSF) through the Sinergia Grants No.~CRSII5 177198/1 and CRSII5\_206008/1, and the Deutsche Forschungsgemeinschaft  (DFG, German Research Foundation) through SFB 1432 (Project-ID 425217212) and Project IDs 449653034 and 510766045.\\
\bibliography{references}
\vspace{1cm}
\appendix
{\bf APPENDIX}

\section{Sample fabrication and vibration detection method}
\label{sec:sample_detection}

\noindent The SiN membranes are fabricated from a 0.5\,mm thick commercial (100) silicon wafer. Both sides of the silicon substrate are coated with $\sim$ 500\,nm thick low-pressure chemical vapor deposited (LPCVD) SiN. The membrane is fabricated on the front side. The backside layer serves as an etch mask. Laser ablation is used to open an etch mask with a typical size of 1.5 $\mathrm{\times}$ 1.5 mm$^{2}$. Using anisotropic etching in aqueous potassium hydroxide (KOH), a hole is etched through the openings of the mask. After the KOH solution reaches the topside layer, the etching stops and a SiN membrane is formed, supported by a massive silicon frame. The membrane is approximately rectangular (542 $\mathrm{\times}$ 524 $\mu$m$^{2}$ in lateral size) in shape, and supports vibrational modes characterized by numbers $(n,m)$ indicating the number of deflection maxima along the $x$- and $y$-directions.

\begin{figure}[t!]
  \centering
  \includegraphics[width=0.6\linewidth]{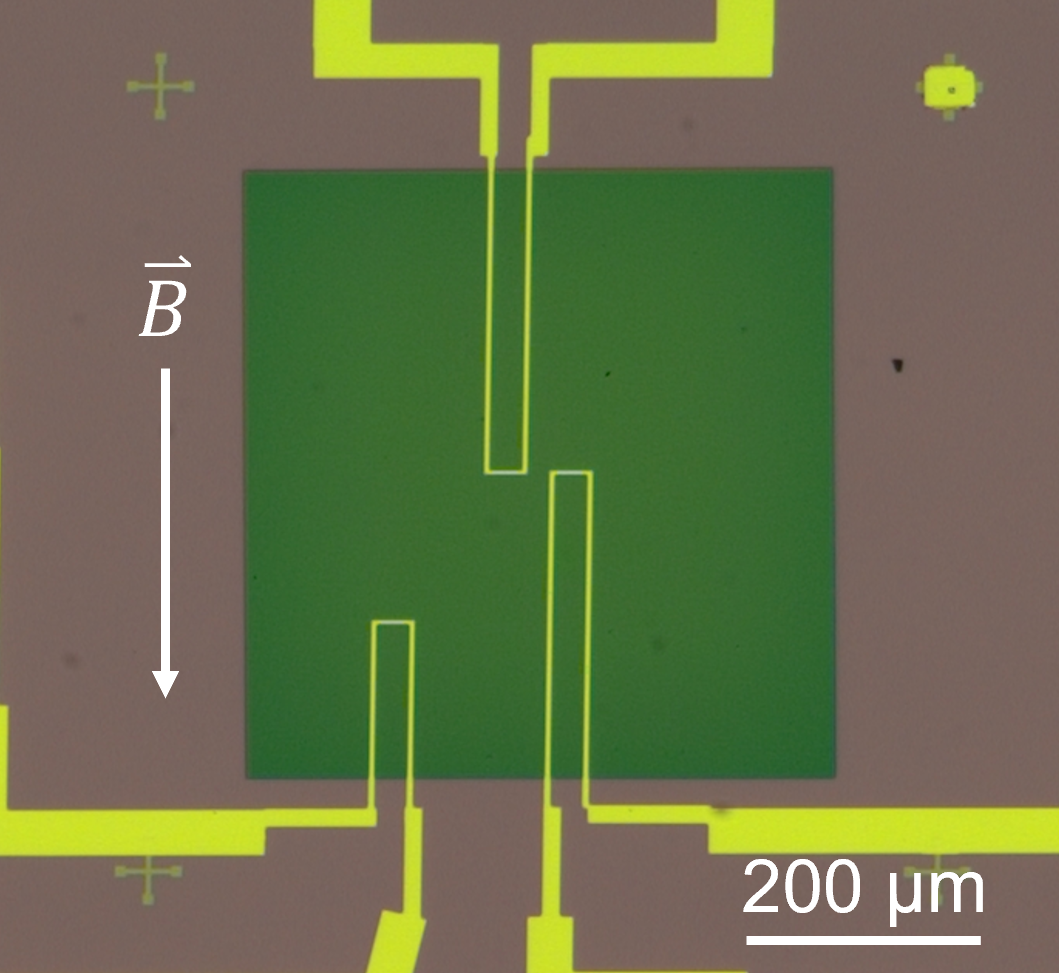}
  \caption{Optical micrograph (top view) of the membrane resonator. Green area: free-standing membrane, brownish area: SiN on top of Si. Yellow lines: three independent electrodes. All measurements reported here have been recorded with the rightmost electrode pair. The arrow indicates the direction of the applied magnetic field.
  }\label{Photo_membrane_probe}
\end{figure}

To detect the mechanical vibration of the SiN membrane, thin aluminum leads ($\sim$ 27 nm) are fabricated on the upper surface of the suspended membranes as well as on the Si frame by standard electron beam lithography and electron beam evaporation. The chip with about $10 \times 10 $ $\textrm{mm}^2$ lateral size carrying the membrane is glued to a piezo disk of 12 mm diameter and 1 mm thickness using a two-component adhesive, see Fig.~\ref{Photo_membrane_probe}.

For the acquisition of the data in the main text, we used the magnetic induction method to characterize the amplitude of the membrane, as depicted in Fig.~1(a) in the main text. The device is placed in a vacuum chamber with a pressure of $p~=~10^{-6}~\textrm{mbar}$ at room temperature and subject to an in-plane magnetic field of flux density amplitude $B$. The Al electrode deposited onto the membrane builds the detection electrodes perpendicular to the magnetic field and their peripheral leads in parallel to the magnetic field. For the sample used in this manuscript, two detection electrodes are located close to the center and in a corner of the membrane, respectively, and both have the length of $L=30~\mu$m.

When the membrane vibrates, the magnetic flux through the area enclosed by the detection electrode and the peripheral leads changes, and thus a potential difference is generated across the electrodes on the membrane. The generated potential difference is first fed to the two input ports of a differential preamplifier to be converted into a single-ended output voltage and to be amplified by a factor ($G_\textrm{diff}$). Then the output voltage ($V_\textrm{out}$) is measured by a lock-in amplifier, a spectral analyzer (SA), and an oscilloscope (OSC). The usage of the differential preamplifier can suppress the common-mode noise (such as the noise generated in the wires and from the vibration of the sample stage) efficiently. The vibration of the membrane part under the peripheral leads does not contribute to the $V_\textrm{out}$ because the peripheral leads are parallel to the magnetic field. Therefore, the vibration velocity ($v$) of the membrane part under the detection electrode is linearly related to $V_\textrm{out}$ by a factor of $(G_\textrm{diff}BL)^{-1}$, where $B=0.45$\,T. Hence, when the membrane is driven by the piezo with the excitation voltage of $V_\textrm{a}$ at the frequency of $\omega_{d}=2\pi f_{d}$, the velocity is $v(t) = Aw_\textrm{d}\cos(w_\textrm{d}t)$ and the real vibration amplitude ($A$) at the position of the detection electrode can be easily calculated by
\begin{equation}
  A = \frac{V_\textrm{out}}{BL\omega_{d}G_\textrm{diff}},
  \label{eq:nm_conver}
\end{equation}
\begin{figure}[htb]
  \includegraphics[width=\linewidth]{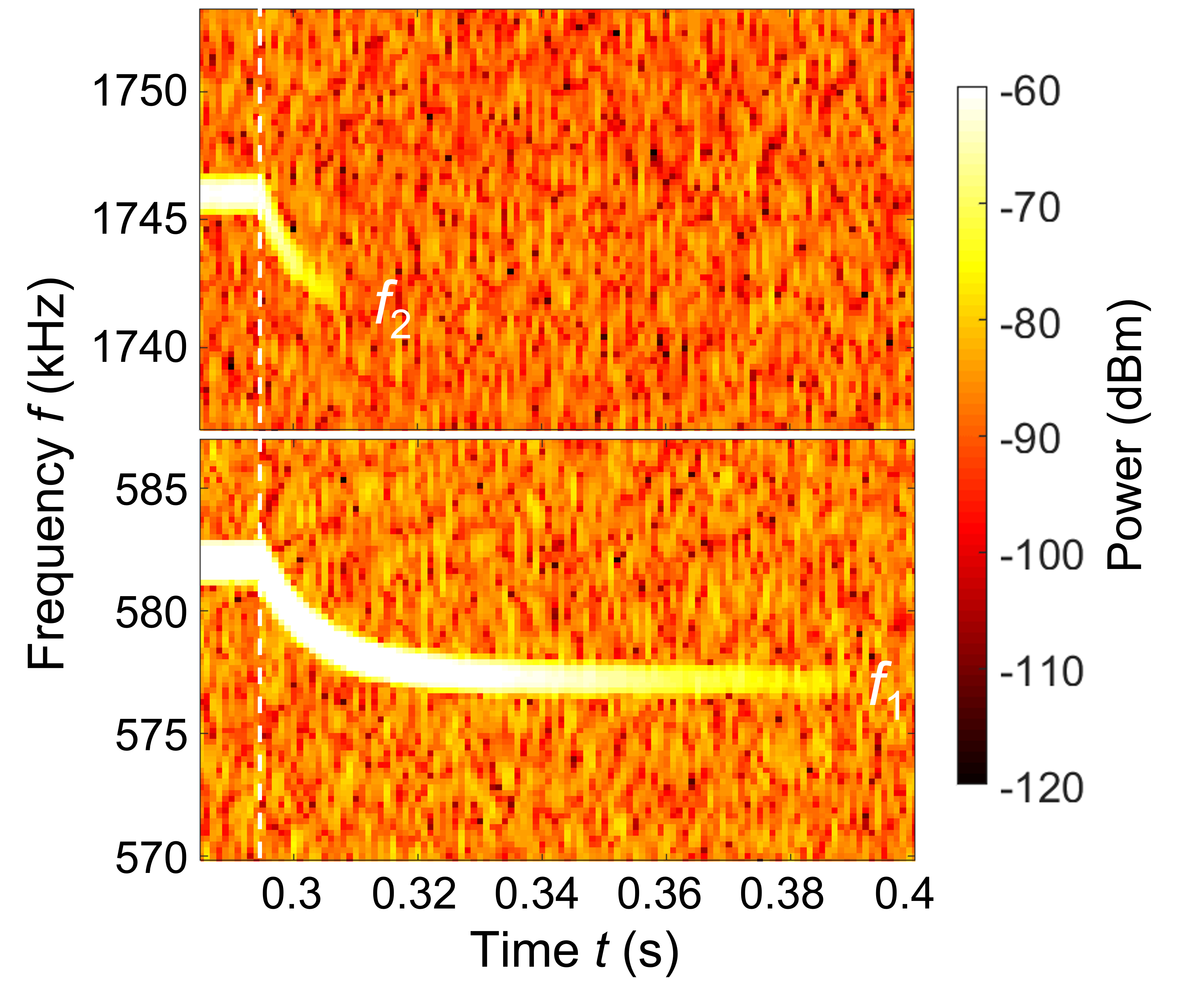}
  \caption{Ringdown measurement under $V_\textrm{a}$ = 100\,mV and $f_d$ = 582\,kHz. The color encodes the intensities of the response in the frequency ranges around $f_d$ and $3f_d$. The drive power is switched off at the dashed vertical line. From the, the amplitude decreases and the frequencies develop toward the eigenfrequencies $f_1$ and $f_2$ corresponding to the $m_1$ and $m_2$ modes, respectively.}.
  \label{fig:ringdown_582kHz}
\end{figure}

\section{Characterization of mechanical properties of membrane}
\label{sec:Mech_properties}

\begin{figure}[h!]
  \centering
  \includegraphics[width=\linewidth]{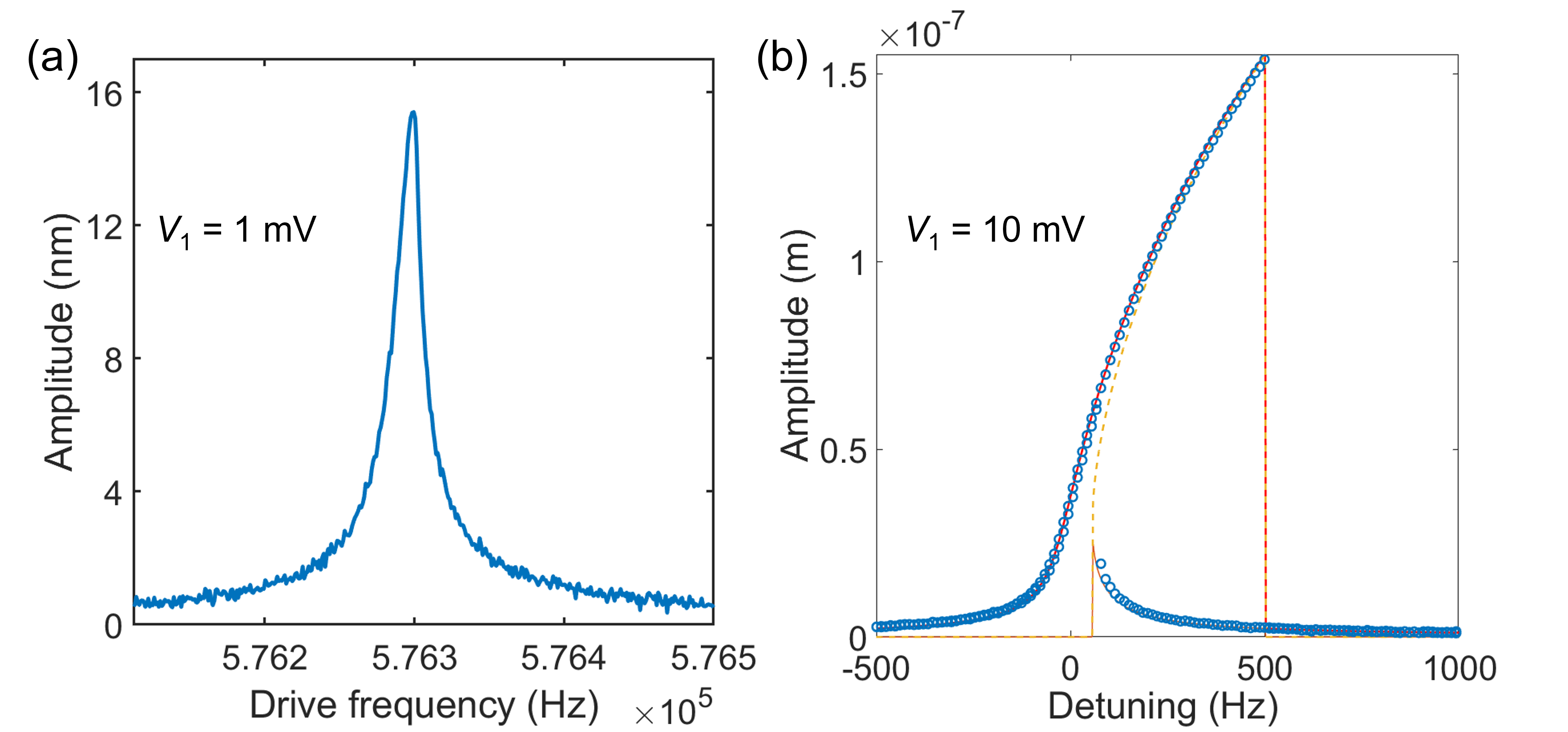}
  \caption{Linear and nonlinear frequency response of $m_1$ of the SiN resonator driven with (a) $V_\textrm{1}$ = 1\,mV and (b) $V_\textrm{1}$ = 10\,mV, respectively. Blue symbols: experimental, ,red/orange line: fit with Duffing model for the upsweep/downsweep.}\label{fig:Duffing_fit_1}
\end{figure}
\begin{figure}[t!]
  \includegraphics[width=\linewidth]{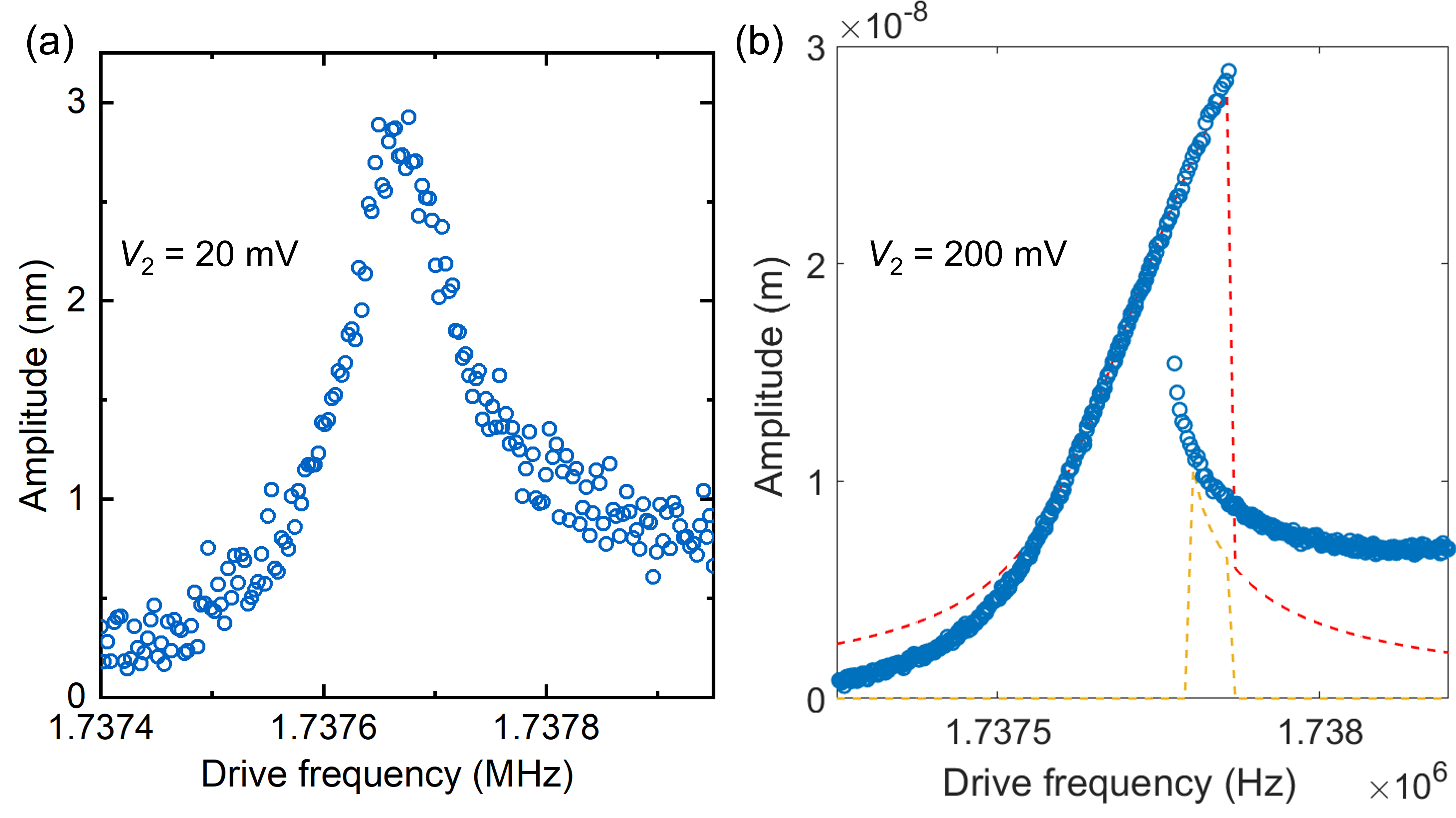}
  \caption{Linear and nonlinear frequency response of the $m_2$ mode driven directly with AC electric voltage (a) $V_\textrm{2}$ = 20\,mV and (b) $V_\textrm{2}$ = 200\,mV, respectively. In (b), the measured amplitudes are plotted as blue dots, and the theoretical calculation of the Duffing model is plotted as red (up sweep) and yellow (down sweep) line. }
  \label{fig:Duffing_fit_2}
\end{figure}
\begin{figure}[t!]
  \centering
  \includegraphics[width=\linewidth]{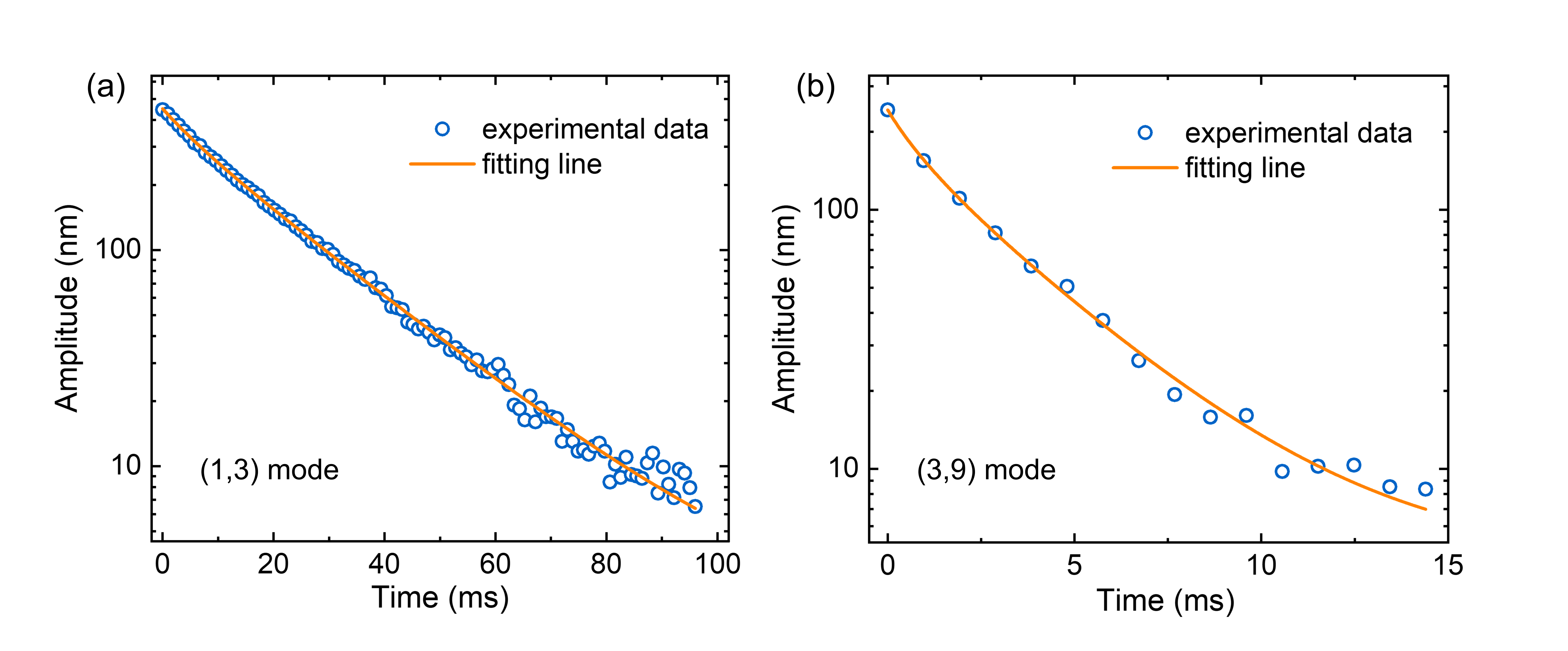}
  \caption{Ringdown measurements of (a)  mode $m_1$ and (b) mode $m_2$. The drive was switched off at $t = 0$. The fits have been calculated with Eq.~\eqref{eq:nonlinear_damping_extraction}.}\label{fig:nonlineardamping_ringdown}
\end{figure}
\noindent The Young's modulus $E$ and the residual stress $\sigma$ of the SiN membrane used in the present work are measured as $E = 213$\,GPa, $\sigma = 0.1$\,GPa, respectively, extracted from the dispersion curve of the bending waves, for details see our previous work~\cite{yang2021mechanically}.

As shown in Fig.~1(b) of the main text, a flexural mode shows response at $f_d$ from 576 to 583 kHz and another flexural mode is coupled into the vibration by 1:3 internal resonance and a shark-fin-shaped frequency response appears at 3$f_d$ ($f_d$ = 581 - 582 kHz). To determine which flexural mode it is, a ringdown measurement is performed. $V_\textrm{a}$ is swept up to 582\,kHz where the amplitude response at 3$f_d$ is large and then $V_\textrm{a}$ is switched off (indicated by the white dashed line in Fig.~\ref{fig:ringdown_582kHz}). The trace of the vibration during the ringdown measurement is captured and converted into frequency spectra by fast Fourier transformation, as shown in Fig.~\ref{fig:ringdown_582kHz}. After the drive power is switched off, the amplitude of the vibration reduces and the frequencies decrease until they finally saturate at the linear eigenfrequencies of these modes, enabling us to identify the modes excited at $f_d$ and 3$f_d$ in Fig.~1(b) as the (1,3)  ($M_1$) and (3,9) ($m_2$) modes, respectively.

We determined the mechanical parameters of the two driven modes by fitting their linear and weakly-nonlinear amplitude responses in Fig.~\ref{fig:Duffing_fit_1} and \ref{fig:Duffing_fit_2} as summarized below:
Eigenfrequencies $f_\textrm{1} \approx 576~\mathrm{kHz}$ and $f_\textrm{2} \approx 1.738~\mathrm{MHz}$, damping rates $\Gamma_\textrm{1}/2\pi = 14~\mathrm{Hz}$ ($Q_\textrm{1} \approx 40600$) and $\Gamma_\textrm{2}/2\pi = 80~\mathrm{Hz}$ ($Q_\textrm{2} \approx 21700$), and Duffing nonlinearities $\beta_\textrm{1} = 1.26 \times 10^{24}~\mathrm{m}^{-2}\mathrm{s}^{-2}$ and $\beta_\textrm{2} = 4.2 \times 10^{25}~\mathrm{m}^{-2}\mathrm{s}^{-2}$ for  $m_1$ and $m_2$, respectively. The extracted effective forces $F_\textrm{1}$ at $V_\textrm{1}$  = 0.1\,V are $500$\,$~\mathrm{m}^{2}$Hz for the (1,3) mode ($m_1$). Note that eigenfrequencies with the same index ratio, e.g., $f_\textrm{1}$ and $f_\textrm{2}$ are not exact multiples from each other because of the small deviation of the membrane from a square shape. In our experiments, temperature drifts have been observed to induce frequency shifts in the order of several hundreds of Hz/K~\cite{yang2023quantitative}. The temperature in the lab may vary by a couple of K. Thus, the absolute values of the eigenfrequencies as well as the drive frequency used to excite the flexural modes may vary by a few kHz from  measurement to measurement.
 
To extract the nonlinear damping coefficients of $m_1$ and $m_2$, we excite one flexural mode into its high-amplitude vibration without coupling other flexural modes into the vibration and then turn off the drive to perform ringdown measurements, as shown in Fig.~\ref{fig:nonlineardamping_ringdown}. In Figs.~\ref{fig:nonlineardamping_ringdown} (a) and (b), we show the nonlinear amplitude decays for $m_1$ and $m_2$, respectively. 
We fit the ringdown curve with~\cite{catalini2021modeling}:
\begin{equation}
  x_j (t) = \frac{x_\textrm{j,0}e^{\frac{-\Gamma_j t}{2}}}{\sqrt{1+\frac{\eta_j}{4\Gamma_j}x_\textrm{j,0}^2(1-e^{-\Gamma_j t})}}\, ,
  \label{eq:nonlinear_damping_extraction}
\end{equation}
Here, $x_\textrm{j,0}$ is the vibration amplitude under the excitation $V_\textrm{j}$; $t$ is the time after turning off $V_\textrm{j}$; $x_j (t)$ is the vibration amplitude as a function of time. The nonlinear damping factors $\eta_j$ of modes $m_1$ and $m_2$ are $8.2 \pm \frac{1}{2} \times 10^{14}$ and $3.8 \pm \frac{1}{2} \times 10^{16}$ Hz/m$^2$, respectively.

\section{Derivation of the model}
\subsection{Rectangular membrane model}
\noindent Here, we derive the equations of motion (1) and (2) of the main text, by describing a stretched rectangular elastic membrane under the influence of an external force~\footnote{This subsection is adapted from Ref.~\cite{kosata2023thesis}}.
We are particularly focused on the out-of-plane displacement \( w \equiv w(x, y, z) \), which is governed by the field Lagrangian~\cite{landau2012theory, ugural2017plates}:
\begin{equation}
  \mathcal{L} = \frac{\rho h}{2} \dot{w}^2 - \frac{\bar{\sigma} h}{2} (\boldsymbol{\nabla} w \cdot \boldsymbol{\nabla} w) - \frac{\alpha}{4} (\boldsymbol{\nabla} w \cdot \boldsymbol{\nabla} w)^2 - f w\,.
\end{equation}
Here, \( \bar{\sigma} \) represents the applied stress, \( \rho \) is the material density, and \( h \) is the membrane thickness. The nonlinearity coefficient, \( \alpha = \frac{E h}{2(1 - \nu^2)} \), depends on Young's modulus \( E \) and Poisson's ratio \( \nu \). Additionally, \( f \equiv f(x, y, t) \) denotes the external force.

Using the Galerkin discretization method~\cite{nayfeh2008nonlinear}, the displacement \( w \) can be represented through the normal modes of the linear system (with \( \alpha = 0 \)). Each mode is described by a pair of integers \( \mathbf{j} = (j_1, j_2) \). For a membrane clamped at the boundaries \( x = 0, L_x \) and \( y = 0, L_y \), the normal modes take the form:
\begin{equation}\label{eq:4.2}
  \phi_{\mathbf{j}}(x, y) = \sin \left( \frac{j_1 \pi x}{L_x} \right) \sin \left( \frac{j_2 \pi y}{L_y} \right).
\end{equation}
These modes serve as a basis to decompose the displacement \( w \), where each mode is factored into its spatial profile \( \phi_{\mathbf{j}}(x, y) \) (dimensionless) and its amplitude \( u_{\mathbf{j}}(t) \) (having dimensions of length):
\begin{equation}
  w(x, y, t) = \sum_{\mathbf{j}} u_{\mathbf{j}}(t) \phi_{\mathbf{j}}(x, y)\, .
\end{equation}
This approach is exact when the sum over \( \mathbf{j} \) extends to infinity. The corresponding Lagrange's equations of motion for each \( u_{\mathbf{j}}(t) \) are:
\begin{multline}\label{eq:4.4}
  \left( \rho h \int_S \phi_{\mathbf{j}}^2 \, dS \right) \ddot{u}_{\mathbf{j}} + \left( \bar{\sigma} h \int_S \nabla \phi_{\mathbf{j}} \cdot \nabla \phi_{\mathbf{j}} \, dS \right) u_{\mathbf{j}} \\
  + \alpha \sum_{\mathbf{k l m}} C_{\mathbf{j k l m}} u_{\mathbf{k}} u_1 u_{\mathbf{m}} = \int_S f \phi_{\mathbf{j}} \dd{S}\, ,
\end{multline}
where \( \int_S dS = \int_0^{L_x} dx \int_0^{L_y} dy \), and \( C_{\mathbf{jklm}} \) is a tensor defined by:
\begin{equation}\label{eq:4.5}
  C_{\mathbf{jklm}} = \int_S dS \left( \boldsymbol{\nabla} \phi_{\mathbf{j}} \cdot \boldsymbol{\nabla} \phi_{\mathbf{k}} \right) \left( \boldsymbol{\nabla} \phi_{\mathbf{l}} \cdot \boldsymbol{\nabla} \phi_{\mathbf{m}} \right)\, .
\end{equation}
The linear part of Eq. \eqref{eq:4.4} provides the natural frequency for each mode:
\begin{equation}
  \omega_{\mathbf{j}} = \sqrt{\frac{\bar{\sigma} \int_S \boldsymbol{\nabla} \phi_{\mathbf{j}} \cdot \boldsymbol{\nabla} \phi_{\mathbf{j}} \, dS}{\rho \int_S \phi_{\mathbf{j}}^2 \, dS}} = \pi \sqrt{\frac{\bar{\sigma}}{\rho} \left( \frac{j_1^2}{L_x^2} + \frac{j_2^2}{L_y^2} \right)}\, .
\end{equation}
The nonlinear part includes terms involving one or more modes, often referred to as self-nonlinearities or cross-nonlinearities. The self-nonlinearity for each mode, expressed as
\begin{equation}
  C_{\mathrm{j} j \mathrm{j} j} = \frac{9\left(j_1 L_x\right)^4 + 2\left(j_1 j_2 L_x L_y\right)^2 + 9\left(j_2 L_y\right)^4}{64\left(L_x L_y\right)^3} > 0\, ,
\end{equation}
introduces a cubic term in \( u_{\mathbf{j}} \) within Eq. \eqref{eq:4.4}. As a result, each mode exhibits cubic nonlinearity, making the system behave as a Duffing (or Kerr) oscillator.

In this particular case, not all potential mode coupling terms are present. By utilizing the analytical expressions for \( \phi_{\mathrm{j}} \), we can identify the non-zero components of the tensor \( C_{\mathrm{jklm}} \). From Eq.~\eqref{eq:4.5} and the mode shapes given in Eq.~\eqref{eq:4.2}. We encounter two spatial integrals, each involving a product of four cosine functions. For \( C_{\mathrm{jklm}} \) to be non-zero, these products must result in a constant contribution in both \( x \) and \( y \). This condition is satisfied if:
\begin{equation}
  j_1 \pm k_1 \pm l_1 \pm m_1 = 0 \quad \text{and} \quad j_2 \pm k_2 \pm l_2 \pm m_2 = 0\, ,
\end{equation}
hold true for at least one combination of the \( \pm \) signs. Evaluating these cross-nonlinearities leads to the equations of motion (1) and (2) in the main text.

\subsection{Slow-flow equations}
\noindent The theoretical model used in the analysis is given by Eq.~(1) and (2) in the main text. We are interested in the stationary responses of both the $m_1$ and $m_2$ modes at the frequency of their respective drives. Therefore, we employ a Floquet expansion in the rotating frame of the drives [cf.~Eq.~(3) and (4) in the main text], separating the fast dynamics from the stroboscopic dynamics. We do this with the help of the open-source software package HarmonicBalance.jl~\cite{kovsata2022harmonicbalance}. Using the package, we can transform the coupled equations from Eqs.~(1) and (2) into slow-flow equations for the quadratures \(u_i\) and \(v_i\):
\begin{widetext}    
\begin{align}
  \label{eq:slow_flow_2_coupled first}
  \dv{u_1}{t}
              & =
  \left(\frac{\Delta_1}{2}
  - \frac{3 \beta_1  }{8\omega_d}X_1^2
  -\frac{\zeta}{4\omega_d} X_2^2
  - \frac{3 \xi }{4\omega_d} Y_{12}
  \right)  v_1
  - \left(\frac{3 \xi }{8\omega_d} X_1^{2}
  + \frac{\zeta}{4\omega_d} Y_{12} \right) v_2
  + \left(\frac{\eta_1}{8}  X_1^{2}- \frac{\gamma_1}{2} \right) u_1\,,
  \\
  \label{eq:slow_flow_2_coupled second}
  \dv{v_1}{t} & =
  \left(\frac{\Delta_1}{2}+\frac{3\beta_1 }{8\omega_d} X_1^2+ \frac{\zeta}{4\omega_d}  X_2^2+ \frac{3  \xi}{4\omega_d} Y_{12}\right)  u_1
  +\left(\frac{3  \xi}{8\omega_d} X_1^2 + \frac{\zeta}{4\omega_d} Y_{12} \right)u_2
  - \left(\frac{ \eta_1 }{8}X_1^{2} +\frac{\gamma_1}{2} \right) v_1
  - \frac{F}{2\omega_d}\,,
  \\
  \dv{u_2}{t} & =
  \left(\frac{\Delta_2}{2} - \frac{\zeta }{12\omega_d} X_1^{2} - \frac{3\beta_2}{24\omega_d} X_2^{2} \right)v_2
  - \frac{\xi}{8\omega_d}Z^u_1  v_1
  + \left(\frac{\eta_2}{8}X_2^2 - \frac{\gamma_2}{2} \right) u_2 \,,
  \\
  \dv{v_2}{t} & =
  \left(\frac{\Delta_2}{2}
  + \frac{\zeta}{12\omega_d}X_1^{2}
  + \frac{3\beta_2}{24\omega_d}X_2^2
  \right) u_2
  + \frac{\xi}{8\omega_d} Z^v_1 u_1
  - \left(\frac{\eta_2 }{8}X_2^2+\frac{\gamma_2}{2}  \right)v_2\,,
  \label{eq:slow_flow_2_coupled last}
\end{align}
\end{widetext}
with detuning $\Delta_1 = \frac{\omega_d^2-\omega_1^2}{2\omega_d}$ and $\Delta_2 = \frac{(3\omega_d)^2-\omega_1^2}{2(3\omega_d)}$, nonlinearity $\beta_i$,  amplitude $X_{i} = (u_i^2+v_i^2)^{1/2}$, linear damping $\gamma_i$, and nonlinear damping $\eta_i$ of the $i$-th resonator. The parameters $\xi$ and $\zeta$ are the coupling coefficients to the $x^3_1x_2$ and $x^2_1x_2^2$ terms, respectively. For convenience we define \( Z^u_i \equiv 3 u_i^2 + v_i^2 \), \( Z^v_i \equiv  u_i^2 + 3 v_i^2 \), and  $Y_{12} \equiv u_1 u_2+ v_1 v_2$. We search for the system's NESS (non-equilibrium stationary states), focusing on the solutions for \(u_i\) and \(v_i\), when \(\dot{u}_i = \dot{v}_i = 0\). Hence, finding the NESS boils down to identifying the roots of the polynomial system defined by Eqs.~\eqref{eq:slow_flow_2_coupled first}-\eqref{eq:slow_flow_2_coupled last}. To achieve this, HarmonicBalance.jl employs Homotopy Continuation~\cite{HomotopyContinuation.jl}, a technique which guarantees to find all the roots and therefore all the NESS of the system.

We obtain a good fit with the measured stationary amplitudes $\sqrt{u_{1}^2+v_{1}^2}$ and $\sqrt{u_{2}^2+v_{2}^2}$, see Fig.~2 of the main text. All the figures are generated using the following parameters: eigenfrequencies $f_\textrm{1} \approx 576.635~\mathrm{kHz}$ and $f_\textrm{2} \approx 1.744375~\mathrm{MHz}$, damping rates $\Gamma_\textrm{1}/2\pi = 14.2~\mathrm{Hz}$ ($Q_\textrm{1} \approx 40600$) and $\Gamma_\textrm{2}/2\pi = 85.5~\mathrm{Hz}$ ($Q_\textrm{2} \approx 20400$), and Duffing nonlinearities $\beta_\textrm{1} = 1.4 \times 10^{24}~\mathrm{m}^{-2}\mathrm{s}^{-2}$ and $\beta_\textrm{2} = 2.8 \times 10^{25}~\mathrm{m}^{-2}\mathrm{s}^{-2}$, nonlinear damping factors $\eta_\textrm{1} = 8.2 \times 10^{14}$ and $\eta_\textrm{2} = 1 \times 10^{14}$ $Hz/m^2$ for $m_1$ and $m_2$, respectively.

\section{Bogoliubov–de Gennes linear response theory}

\subsection{Jacobian}
\noindent In the main text, we derived the linear response op top of the NESS to theoretically interpret the PNP measurements. This was done by linearizing the equations of motion for the effective system, \eqref{eq:slow_flow_2_coupled first}-\eqref{eq:slow_flow_2_coupled last}, around the rotating NESS \( \vb{u}^* \). We introduce a small perturbation \( \delta \vb{u} = \vb{u} - \vb{u}^* \), which leads to a new set of differential equations:
\begin{equation} \label{eq:linEOM}
  \dv{t} \delta \vb{u} = \mathcal{J}(\vb{u}^*) \delta \vb{u}\, ,
\end{equation}
where the dynamics of the perturbations are governed by the eigensystem of the \( 4 \times 4 \) Jacobian \( \mathcal{J} \), evaluated at \( \vb{u}^* \). The Jacobian is expressed as:
\begin{widetext}    
\begin{equation} \label{eq:Jac}
  \mathcal{J} = J + \mqty(
  -\frac{\gamma_1}{2} - \frac{\eta_1}{8} Z_1^u & - \frac{\eta_1}{8} u_1v_1 & 0&0\\
  - \frac{\eta_1}{8} u_1v_1 & -\frac{\gamma_1}{2} - \frac{\eta_1}{8} Z_1^v & 0&0\\
  0&0&-\frac{\gamma_2}{2} - \frac{\eta_2}{8} Z_2^u & - \frac{\eta_2}{8} u_2v_2 \\
  0&0&- \frac{\eta_2}{8} u_2v_2 & -\frac{\gamma_2}{2} - \frac{\eta_2}{8} Z_2^v
  )\, ,
\end{equation}
with
\begin{equation}
  J = \left(
  \begin{array}{cccc}
    \frac{3 (\xi u_1 v_2-\xi u_2 v_1+\beta_1 u_1 v_1)}{4 \omega_d }                    & \frac{-6 \xi Y_{12}+2 \zeta X_2^2+3\beta_1 Z^v_1}{8 \omega_d } +\frac{\Delta_1}{2} & \frac{2 \zeta u_2 v_1-3 \xi u_1 v_1}{4 \omega_d }                       & \frac{3 \xi (u_1^2-v_1^2)+4 \zeta v_1 v_2}{8 \omega_d }                \\
    -\frac{6 \xi Y_{12}+2 \zeta X_2^2+3\beta_1 Z^u_1}{8 \omega_d } +\frac{\Delta_1}{2} & -\frac{3 (\xi u_1 v_2-\xi u_2 v_1+\beta_1 u_1 v_1)}{4 \omega_d }                   & -\frac{3 \xi u_1^2-3 \xi v_1^2+4 \zeta u_1 u_2}{8 \omega_d }            & -\frac{u_1 (3 \xi v_1+2 \zeta v_2)}{4 \omega_d }                       \\
    \frac{u_1 (3 \xi v_1+2 \zeta v_2)}{12 \omega_d }                                   & \frac{3 \xi (u_1^2-v_1^2)+4 \zeta v_1 v_2}{24 \omega_d }                           & \frac{\beta_2 u_2 v_2}{4 \omega_d }                                     & \frac{2 \zeta X_1^2+3\beta_2 Z^v_2}{24 \omega_d } + \frac{\Delta_2}{2} \\
    -\frac{3 \xi u_1^2-3 \xi v_1^2+4 \zeta u_1 u_2}{24 \omega_d }                      & \frac{v_1 (3 \xi u_1-2 \zeta u_2)}{12 \omega_d }                                   & -\frac{2 \zeta X_1^2+3\beta_2 Z^u_2}{24 \omega_d } + \frac{\Delta_2}{2} & -\frac{\beta_2 u_2 v_2}{4 \omega_d }                                   \\
  \end{array}
  \right).
\end{equation}
\end{widetext}
Given that Eq.~\eqref{eq:linEOM} is a set of 1st order ordinary differential equations, the system’s dynamics are then determined by the exponentials \( e^{\lambda_k t} \), where \( \lambda_k \) are the eigenvalues of the Jacobian.  If \( \mathrm{Re}(\lambda_k) < 0 \) for all \( \lambda_k \), the NESS \( \vb{u}^* \) is stable. However, if \( \mathrm{Re}(\lambda_k) > 0 \) for at least one eigenvalue, the state becomes unstable, and perturbations like noise or small external drives will push the system away from \( \vb{u}^* \). The eigenvalues for the NESS for both the upward and downward sweeps are shown in Fig. 2 of the main text.

\subsection{Linear response in the rotating frame}
\noindent To determine the linear response of an NESS to an additional oscillatory force, such as weak probes or noise, we solve for the perturbation \(\delta \mathbf{u} \) in the presence of an external drive \(\boldsymbol{\sigma} \, e^{i \Omega t} \), where $\Omega$ is the probe frequency. This is captured by the equation:
\begin{equation} \label{eq:DrivenLinEOM}
  \dv{t} \delta \mathbf{u} =  \mathcal{J}(\vb{u}^*) \delta \mathbf{u} + \boldsymbol{\sigma} \, e^{i \Omega t}\, .
\end{equation}
In order to solve this equation, we first assume the form \( \delta \mathbf{u}_k = A_k(\Omega) \mathbf{w}_k e^{i \Omega t} \), where the eigenvector \( \mathbf{w}_k \) is such that \( \mathcal{J}(\mathbf{u}^*) \mathbf{w}_k = \lambda_k \mathbf{w}_k \), with \( \lambda_k \) serving as the corresponding eigenvalue. By calculating the amplitude \( A_k(\Omega) \) of the response, we arrive at the following:
\begin{equation}
  A_k(\Omega) = \frac{\boldsymbol{\sigma} \cdot \mathbf{w}_k}{-\operatorname{Re}(\lambda_k) + i \left( \Omega - \operatorname{Im}(\lambda_k) \right)}\, .
\end{equation}
 This shows that each eigenvalue \( \lambda_k \) produces a linear response that follows a Lorentzian profile, centered at \( \Omega = \operatorname{Im}(\lambda_k) \). Essentially, the system’s response resembles that of a harmonic oscillator, with a resonance frequency \( \operatorname{Im}(\lambda_k) \) and damping \( \operatorname{Re}(\lambda_k) \).
 
\subsection{Excitation on top of the NESS}
\noindent In many fields, the response described above are thought of as quasiparticle living in the linearized potential on top of the stationary state \(\mathbf{u}^*\) with a characteristic frequency \( \operatorname{Im}(\lambda_k) \) and lifetime \( \operatorname{Re}(\lambda_k) \). Indeed, in fields such as superconducting circuits or quantum optics, the above procedure is called the Bogoliubov-de-Gennes formalism~\cite{xiao2009}. In this language, the NESS $\vb{u}^*$ acts as a chemical potential for the fluctuation, defining the local minima of the rotating potential landscape~\cite{Soriente_2020}. When a quasiparticle is excited with an energy cost \( \operatorname{Im}(\lambda_k) \), it leaves behind a quasihole with the opposite energy. Indeed, because the Jacobian \eqref{eq:Jac} is real, the eigenvalues of the Jacobian appear in two complex conjugate pairs, resembling a particle at \( \operatorname{Im}(\lambda_k) \) on a hole at -\( \operatorname{Im}(\lambda_k) \). During a probe experiment in this rotating picture both modes get excited manifesting as sidebands, c.f. Ref.~\cite{Heugel2022TheRO}. How such two sidebands manifest in our PNP experiment in the lab frame is discussed in Sec.~\ref{sec: LR lab}.

Which of the two complex conjugate sidebands is the absorbing or emitting excitation is determined by the symplectic norm~\cite{dumont2024hamiltonian,Soriente_2020,Soriente_2021}:
\begin{align}
  \dd{S}^2 &\equiv \sum\dd{S}_j^2=  i\sum_j\left[c^{u_j}_{k} (c^{v_j}_{k})^*-(c^{u_j}_{k})^* c^{v_j}_{k}\right]\,,\\
  &= 2\sum_j (\Re(c^{u_j}_k) \Im(c^{v_j}_k) - \Im(c^{u_j}_k) \Re(c^{v_j}_k))\,,
\end{align}
where \(c^{u_j}_k\) and \(c{^v_j}_k\) are the components of the eigenvector \(\mathbf{w}_k=(c^{u_1}_{k},c^{v_1}_{k},c^{u_2}_{k},c^{v_2}_{k})\) of the Jacobian \eqref{eq:Jac}. Here, we assume that the eigenvectors are euclidean normalised, i.e., $||\mathbf{w}_k||_2=\sqrt{\mathbf{w}_k\cdot\mathbf{w}_k}=1$. For a given eigenvector \(\mathbf{w}_k\), it can be positive or negative, indicating whether the excitation is particle-like (\(\dd{S}^2 > 0\)) or hole-like (\(\dd{S}^2 < 0\)). Here, it is valid to sum the symplectic norm of the individual rotating frame subspaces $\dd{S}_j^2 $ as long as the modes are not hybridised.

Naively, one would think that the excitation with a positive $\Im(\lambda_k)$ would be the particle-like excitation, and the one with a negative $\Im(\lambda_k)$ would be the hole-like excitation. However, as the excitation is on top of an attractor in a rotating frame, potential maxima can also be an NESS~\cite{Soriente_2020}. Having a stationary state on top of a local maxima results in a reverse excitation spectrum. The symplectic norm exactly measures the local ``curvature'' of a Hamiltonian defined on a symplectic manifold~\cite{dumont2024hamiltonian}. The hole-like excitation exactly reflects that a shift in negative frequency from the drive $\omega_{i,d}$ is needed to be resonant with the natural frequency $\Omega_i$ in the linear regime, i.e., the stationary state is blue detuned ($\omega_{i,d}>\Omega_i$)~\cite{Soriente_2021,villa2024MStop}.

It might be easier to think about this in a quantum driven-dissipative formalism. Here, however, we stay in classical mean-field limit, but borrow the quantum language. As was shown in the supplemental material of Ref.~\cite{seibold2024floquetexpansioncountingpump}, The quadratures $u_i$ and $v_i$ of mode $m_i$ can be re-expressed as a complex variable $\alpha_i$ and its complex conjugate $\alpha_i^*$ via the transformation:
\begin{equation}
  \mqty(u_j\\ v_j)=\mathbf{S}^{-1}\mqty(\alpha_j\\ \alpha^*_j)=\sqrt{\frac{\hbar}{2 \omega_{d,j}}}\left(\begin{array}{cc}
      1 & 1  \\
      i & -i
    \end{array}\right)\mqty(\alpha_j\\ \alpha^*_j)\,,
\end{equation}
with $\omega_{d,j}$ the frequency ay which the variables are rotated at [cf.~Eqs.~(1)~and~(2)]. Therefore, we have coherent state $\alpha_j=u_j -i v_j$. This change of basis can be seen as a transition to the mean-field limit of a quantum harmonic oscillator's bosonic creation and annihilation operators. In this context, \(\alpha_i = \langle \hat{a}_i \rangle\), where \(\hat{a}_i\) is the bosonic annihilation operator and \(\langle \cdots \rangle\) denotes the expectation value. Note, that we have quantised in the frequency of multiplied of the driving frquency. This is needed to ensure the quantum to classical limit in Floquet theory~\cite{seibold2024floquetexpansioncountingpump}.

Focusing on the subspace of the individual rotating frames, we define $\vb{w}_{k,j}\equiv(c^{u_j}_k,c^{v_j}_k)$. We can rewrite the symplectic norm in term of the eigenvectors in the new basis \mbox{$\vb{v}_{k,j}=S\vb{w}_{k,j}$}, by
$$
\begin{aligned}
  \dd{S}^2_j  & =
\vb{w}_{k,j}^\dagger
\mqty(0& -i\\ i & 0)
\vb{w}_{k,j}, \\
& =\vb{w}_{k,j}^\dagger
\mathbf{S}^{\dagger}\left(\mathbf{S}^{\dagger}\right)^{-1}
\mqty(0& -i\\ i & 0)
\mathbf{S}^{-1} \mathbf{S}\,
\vb{w}_{k,j}\,,\\
& =\left(\mathbf{S} \vb{w}_{k,j}\right)^{\dagger}
\left(\mathbf{S}^{\dagger}\right)^{-1}
\mqty(0& -i\\ i & 0)\mathbf{S}^{-1}
 \mathbf{S} \vb{w}_{k,j}\,, \\
& =\vb{v}_{k,j}^{\dagger} \mathbf{I}_{-} \vb{v}_{k,j}\,,
\end{aligned}
$$
with $\mathbf{I}_{-}=\operatorname{diag}(1,-1)$. Writing this out in term of the original coefficient, we find
$$
 \dd{S}^2_j = (c^{u_j}_k-ic^{v_j}_k)^2 - (c^{u_j}_k+ic^{v_j}_k)^2.
$$
This shows that the norm measures the hole vs particle nature by comparing their relative weights.

\subsection{Linear response in the lab frame}\label{sec: LR lab}

\noindent The PNP response in Fig. 2 of the main text is measured in the lab frame. Hence, we need to determine the perturbation in terms of the ``natural'' variables \(x_i(t) \). To achieve this, we re-express the solution found for Eq.~\eqref{eq:DrivenLinEOM} as $\delta \mathbf{u}_k = A_k(\Omega) \left(\mathbf{w}_k \, e^{i \Omega t} + \mathbf{w}_k^* \, e^{-i \Omega t} \right)$.
Substituting this into the ansatz from Eqs. (3) and (4) of the main text, and reducing the trigonometric terms, we obtain:
\begin{align} \label{eq:deltax}
  \delta x_{i,k}(t) = & \left( \operatorname{Re}(c^{u_i}_k) - \operatorname{Im}(c^{v_i}_k) \right) \,  \cos((\omega_{d,i} - \Omega) t)\nonumber\\
  &+ \left( \operatorname{Im}(c^{u_i}_k) + \operatorname{Re}(c^{v_i}_k) \right) \,\sin((\omega_{d,i} - \Omega) t)  \nonumber            \\
                      & + \left( \operatorname{Re}(c^{u_i}_k) + \operatorname{Im}(c^{v_i}_k) \right) \,\cos((\omega_{d,i} + \Omega) t)\nonumber\\
  &+ \left( \operatorname{Re}(c^{v_i}_k) -\operatorname{Im}[c^{u_i}_k] \right) \,\sin((\omega_{d,i} + \Omega) t)
\end{align}
where \( c^{u_i}_k \) and \( c^{v_i}_k \) represent the components of \( \delta \mathbf{u} \) corresponding to the harmonics \( \omega_{d,i} \). This shows that a motion of the harmonic variables at frequency \( \Omega \) results in motion of \( \delta x_i(t) \) at frequencies \( \omega_{d,i} \pm \Omega \).

Assuming the the vector \( \delta \mathbf{u} \) to be normalized, we define the Lorentzian distribution:
\begin{equation}
  L(x)_{x_0, \gamma} = \frac{1}{(x - x_0)^2 + \gamma^2},
\end{equation}
making all components of \( \delta x_{i,k}(t) \) [from~Eq.~\eqref{eq:deltax}] proportional to \( L(\Omega)_{\operatorname{Im}[\lambda], \operatorname{Re}[\lambda]} \). With the definition of the Lorentzian distribution, we can express the linear response function in Fourier space, \( \chi(\tilde{\omega}) \), as follows:
\begin{align}
  |\chi [\delta x_i](\tilde{\omega})|^2 =& \left( 1 + \alpha_{i,j} \right) L(\tilde{\omega})_{\omega_{d,i} - \operatorname{Im}[\lambda], \operatorname{Re}[\lambda]} \\
  &+ \left( 1 - \alpha_{i,j} \right) L(\tilde{\omega})_{\omega_{d,i} + \operatorname{Im}[\lambda], \operatorname{Re}[\lambda]},
\end{align}
where we used the fact that \( L(x)_{x_0, \gamma} = L(x + \Delta)_{x_0 + \Delta, \gamma} \), and defined:
\begin{equation}
  \alpha_{i,k} =2 \left(\operatorname{Re}[c^{u_i}_k] \operatorname{Im}[c^{v_i}_k]- \operatorname{Im}[c^{u_i}_k] \operatorname{Re}[c^{v_i}_k] ] \right).
\end{equation}
This solution holds for each eigenvalue \( \lambda_k \) of the Jacobian. The linear response function \( \chi[\delta x_{i,k}](\tilde{\omega}) \) for each eigenvalue \( \lambda_k \) and harmonic \( \omega_{d,i} \) consists of:
\begin{itemize}
  \item A Lorentzian centered at \( \omega_{d,i} - \operatorname{Im}[\lambda_k] \), with amplitude \( 1 + \alpha_{i,k} \).
  \item A Lorentzian centered at \( \omega_{d,i} + \operatorname{Im}[\lambda_k] \), with amplitude \( 1 - \alpha_{i,k}\).
\end{itemize}
Thus, the linear response of the system in the state \( \vb{u}^* \) is fully characterized by the complex eigenvalues and eigenvectors of \( \mathcal{J}(\vb{u}^*) \).

\section{Classification of noise sidebands to the excited flexural modes} 
\label{sec:classification_sidebands}
\noindent Experimentally, the eigenmodes of a flexural mode $j$ can be visualized by the system fluctuations around the NESS ($f$ = $f_d$) as noise sidebands~\cite{huber2020spectral, yang2021mechanically}. Here, to distinguish the eigenvalue of the mode $j$ in linear and nonlinear resonator, we denote it as eigenmode in the linear resonator and linear response in the nonlinear resonator which is generated by a strong drive power. The frequencies of the noise sidebands strongly depend on the vibration amplitude, nonlinearity and the detuning frequency of the mode~\cite{huber2020spectral, yang2021mechanically}. When the vibration amplitude of the mode is low, the noise sideband has the same or similar frequency as the natural frequency $f_j$, and only one noise sideband can be well observed. Instead, when the vibration amplitude of the mode is large, especially entering into the upper branches with relatively large detuning, one eigenmode splits into two linear responses and thus a pair of noise sidebands can be observed. In this regime, the frequency spacing between the pair of noise sidebands and $f_d$ becomes relatively constant~\cite{Steele2021ACStark}.

Independently of the detailed modeling explained in the main text, we can also assign the sidebands to their respective modes from a purely experimental point of view, The frequency spacing  of the outer pair of sidebands at $f_d$ = 581.7\,kHz) hardly changes (within 100 Hz) with varying $f_d$ in the detuning range shown in Fig.~2 in the main text. Comparing with the amplitude response of the two coupled modes in Fig.~1(b) in the main text, this pair of sidebands is assigned to $m_1$ because the vibration amplitude and detuning of $m_1$ are large. In contrast, the inner pair of the sidebands  at $f_d$ = 581.7\,kHz presents a significant $f_d$ dependence of the frequency spacing at $f_d$ = 581 - 582 kHz. Therefore, they correspond to $m_2$.

In addition, we can also distinguish the linear responses of $m_1$ and $m_2$ by the bandwidth of the noise sidebands. The linear response of a flexural mode vibrating in the nonlinear regime shows a similar bandwidth as the eigenmode~\cite{huber2020spectral}. Therefore, the linear response of $m_2$ shows a higher damping rate, and the noise sidebands are spectrally broader than those of $m_1$. According to the features described above, the comparison of the frequency spectra in Fig.~2 (b) with the frequency sweep in Fig.~1 (b) confirms the assignment.

\section{2D frequency spectra under different drive power}
\label{sec:80mV_2Dspectra}
\begin{figure}[h]
  \includegraphics[width=1.0\linewidth]{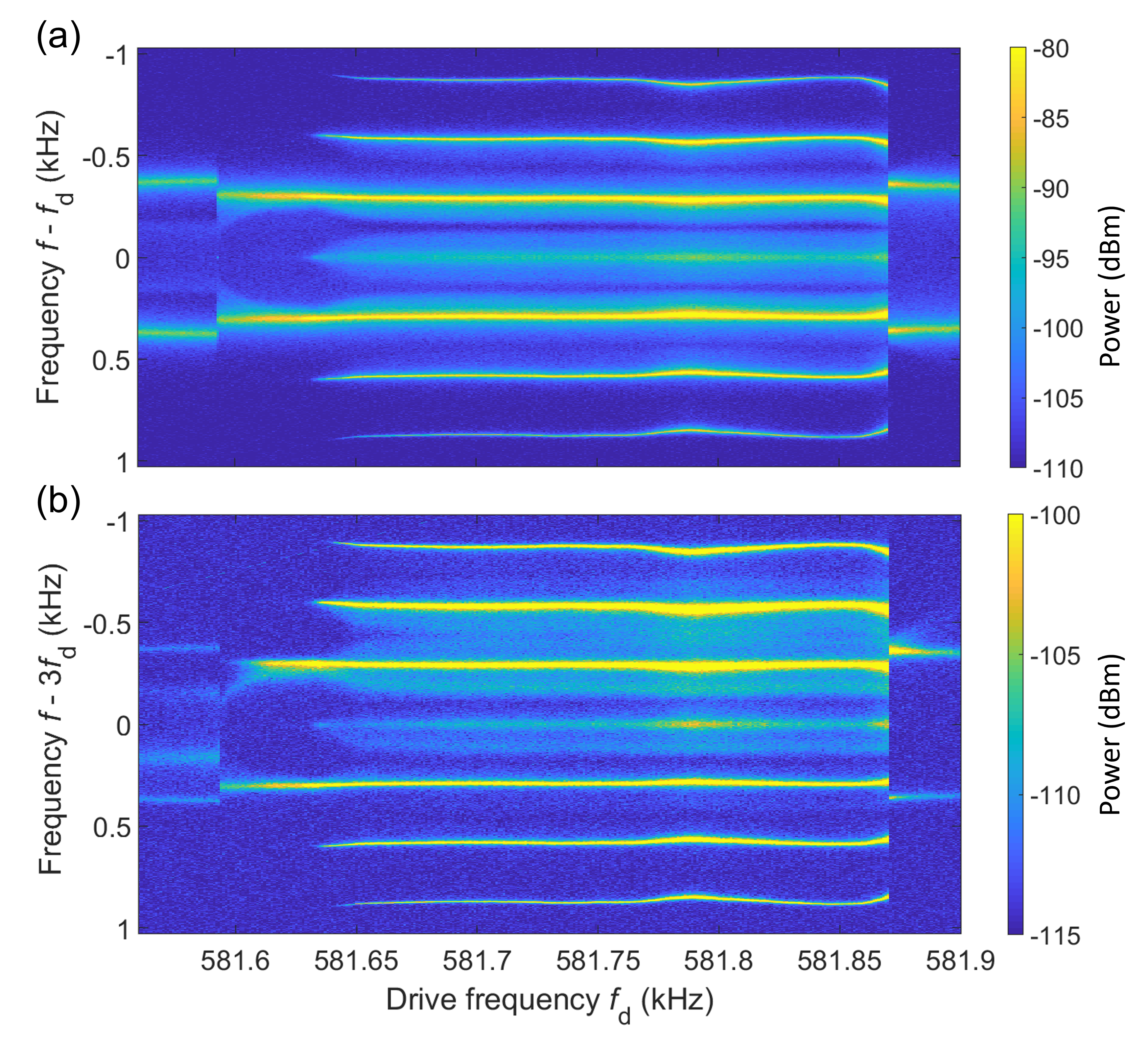}
  \caption{2D frequency spectra of the frequency comb excited under $V_\textrm{1}$ = 80\,mV. (a) and (b) show the spectra around $f_d$ and 3$f_d$, respectively.
  }
  \label{fig:comb-80mV}
\end{figure}
\noindent The generation of the frequency comb has been monitored for different drive powers. Fig.~\ref{fig:comb-80mV} shows the 2D frequency spectra of the mechanical vibration under $V_\textrm{1}$ = 80\,mV with a PNP measurement. Before and after the frequency comb is excited, two sets of noise sidebands are observed. As in the case of $V_\textrm{1}$ = 100\,mV, the frequency comb is excited when the frequencies of these two sets of noise sidebands approach each other. The frequency spacing between the sidebands of the frequency comb is smaller than that observed under $V_\textrm{1}$ = 100\,mV.

\end{document}